\documentclass[11pt,twoside]{article}

\usepackage{asp2006}

\usepackage{natbib}

\def\deg{\hbox{$^\circ$}}
\def\sun{\hbox{$\odot$}}

\def\lesssim{\mathrel{\hbox{\rlap{\hbox{\lower4pt\hbox{$\sim$}}}\hbox{$<$}}}}
\def\gtrsim{\mathrel{\hbox{\rlap{\hbox{\lower4pt\hbox{$\sim$}}}\hbox{$>$}}}}
\def\la{\mathrel{\hbox{\rlap{\hbox{\lower4pt\hbox{$\sim$}}}\hbox{$<$}}}}
\def\ga{\mathrel{\hbox{\rlap{\hbox{\lower4pt\hbox{$\sim$}}}\hbox{$>$}}}}

\def\farcm{\hbox{$.\mkern-4mu^\prime$}}
\def\farcs{\hbox{$.\!\!^{\prime\prime}$}}
\def\fp{\hbox{$.\!\!^{\scriptscriptstyle\rm p}$}}

\def\nodata{\multicolumn{1}{c}{$\cdots$}}
\def\ion#1#2{#1$\;${\small\rm\@Roman{#2}}\relax}

\def\farcm@mss{\mbox{$.\mkern-4mu^\prime$}}%
\def\farcm\farcm@mss 
\def\farcs@mss{\mbox{$.\!\!^{\prime\prime}$}}%
\def\farcs\farcs@mss 
\def\fp{\mbox{$.\!\!^{\scriptscriptstyle\mathrm p}$}}%
\def\farcm@apj{%
 \mbox{.\kern -0.7ex\raisebox{.9ex}{\scriptsize$\prime$}}%
}%
\def\farcs@apj{%
 \mbox{%
  \kern  0.13ex.%
  \kern -0.95ex\raisebox{.9ex}{\scriptsize$\prime\prime$}%
  \kern -0.1ex%
 }%
}%
\def\ion#1#2{#1$\;${\small\rmfamily\@Roman{#2}}\relax}%
\def\nodata{ ~$\cdots$~ }%
\def\la=\lesssim            
\def\ga=\gtrsim 

\def\Sun{\sun}%
\def\sun{\odot}%
\def\aj{AJ}%
\def\araa{ARA\&A}%
\def\apj{ApJ}%
\def\apjl{ApJ}%
\def\apjs{ApJS}%
\def\ao{Appl.~Opt.}%
\def\apss{Ap\&SS}%
\def\aap{A\&A}%
\def\mnras{MNRAS}%
\def\pasp{PASP}%
\def\solphys{Sol.~Phys.}%
\def\nat{Nature}%
\def\gca{Geochim.~Cosmochim.~Acta}%
\def\grl{Geophys.~Res.~Lett.}%
\def\planss{Planet.~Space~Sci.}%
\def\astap=\aap 
\def\apjlett=\apjl 
\def\apjsupp=\apjs 
\def\applopt=\ao 

\def\nodata{ ~$\cdots$~ }%
\def\ulap#1{\vbox\@to\z@{{\vss#1}}}%
\def\dlap#1{\vbox\@to\z@{{#1\vss}}}%

\begin{document}
\title{Expanding and Improving the Search for Habitable Worlds}   
\author{Avi M. Mandell\footnotemark}  
\affil{NASA GSFC}    
\footnotetext{NASA Post-doctoral Fellow}

\begin{abstract} 
This review focuses on recent results in advancing our understanding of the location and distribution of habitable exo-Earth environments.  We first review the qualities that define a habitable planet/moon environment.  We extend these concepts to potentially habitable environments in our own Solar System and the current and future searches for biomarkers there, focusing on the primary targets for future exploratory missions:  Mars, Europa, and Enceladus.  We examine our current knowledge on the types of planetary systems amenable to the formation of habitable planets, and review the current state of searches for extra-solar habitable planets as well as expected future improvements in sensitivity and preparations for the remote detection of the signatures of life outside our Solar System.
\end{abstract}

\section{Introduction} 
We currently have concrete evidence of life on only one planet in the Universe: Earth.  Over the last decade we have taken great strides in the quest to expand this tally, both through the investigation of other planets and moons beyond Earth as well as the investigation of exotic environments on Earth as analogs to potentially habitable environments elsewhere. However, the broad interdisciplinary research field known as ``astrobiology", roughly defined to include any investigation that expands our understanding of the origin, evolution and distribution of life in the universe,  encompasses such a vast range of research topics and core scientific disciplines that it cannot be adequately covered in a single review; in fact, only a brief sampling of research in a few core disciplines will be covered here.

This review will discuss the status of current research in areas of astrobiology related to astronomy and planetary science, focusing on improvements in our understanding of the constraints on the locations of potentially habitable environments as well as current and future efforts to detect life in our own Solar System and beyond.  In \S2 I introduce current theory on the constraints that define the term ``habitable", and in \S3 I discuss the potential for habitable planetary environments elsewhere in our own Solar System.  We must then place our own planetary system in a wider context; in \S4 I present the current paradigms and conflicts in our understanding of planet formation, and specifically in the origin of habitable planets, and I present a review of current and future searches for habitable, and inhabited, environments outside our own Solar System.

\section{Factors in Assessing Planetary Habitability}
The standard definition for a habitable planet has traditionally been one that can sustain 
life similar to that on Earth on its surface or subsurface for a significant period of time. However, this definition is based on our understanding of the current locales for life on Earth and on our current understanding  of the environments present on other planets; it is therefore constantly being redefined as we hypothesize or discover new environments in which life can sustain itself. 

The main requirements for life as we know it are:
\begin{itemize}
\item The presence and stability of liquid water over long time periods
\item The availability of the basic organic building blocks of life (CHNOPS, building blocks, and nutrients)
\item The availability of energy for assembly of biological structures and metabolic processes
\end{itemize}

As we evaluate potentially habitable environments we must keep these fundamental requirements in mind. Also, to decrease confusion with respect to the characteristics of rocky bodies in various environments, in this section the word ``planet" will be used to signify any large rocky body, either orbiting the central star (a traditional ``terrestrial planet") or orbiting another large body in the system (traditionally called a ``satellite" or ``moon").
 
\subsection{Characteristics Affecting Surface and Subsurface Habitability} 
The fundamental requirement for liquid water is a clement mean 
environmental temperature.  Temperatures between 0\deg C (273 K) and 100\deg C (373 K) are 
necessary for pure water to form a liquid at standard temperature and pressure, but 
pressure and solutes can dramatically change these limits; extant life has been detected 
in water between -20\deg C (253 K)  and 121\deg C (394 K) \citep{rivkina00,kashefi03}. On rocky planets or moons this environment can be either on or below the surface, depending on the incident radiation and the subsurface heat source; the temperature of a planetary environment is a function of the balance between heating and radiation into space.  This balance is affected by both surface processes and the bulk characteristics of the planet itself.
 
\subsubsection{Surface Temperature} 
For surface environments, the temperature is modulated primarily by the ratio of radiation absorbed from the central star (or stars) to the radiation radiated to space by the planet's surface and atmosphere. Stellar heat input will be sensitive to the size of the central star and the orbital distance. We can estimate a ``Habitable ZoneÓ or range in orbital semi-major axis for a given stellar type where water can exist as a liquid \citep{hart78}; for a G-type star such as our Sun, the traditional Habitable Zone lies between 0.95 AU and 1.37 AU \citep{kasting93}. The surface temperature can be increased if the incident radiation can be retained by the planet's atmosphere (known as the ``greenhouse effect"); the potential for heat retention depends on the composition of the atmosphere, and for thick atmospheres of molecules such as carbon dioxide and methane the surface temperature may be raised significantly. The greenhouse effect may therefore extend the habitable zone out to 2.4 AU for G-type stars \citep{forget97}. 
 
\subsubsection{Subsurface Temperature}
In subsurface environments the local temperature is defined by heat transfer through the planet's interior. The heat source is either internal latent heat from accretion (as on Earth) or an external force such as tidal compression (as on Io), while cooling is limited by insulation from surface layers. The decay of radioactive isotopes also makes a small contribution, and may be enough to maintain liquid water in a subsurface layer in the absence of additional heat sources (a possibility for Enceladus; \citet{schubert07}). Heat can also be generated by tidal forces in moons orbiting a giant planet. As the moon travels closer to and farther from the parent planet, the change in gravity causes it to expand towards the giant planet. The internal structure of the moon continually compresses and expands, creating frictional heating that can be conducted throughout the moon. This heat may also be sufficient to maintain mantle convection and subsurface liquid water (a possibility on Europa and/or Enceladus; see \S3.2).  

\subsubsection{Planet Mass and Structure} 
The internal structure of a planet will have a direct impact on the stability of its climate. On 
Earth, internal heating mechanisms result in plate tectonics and volcanism, both of which 
play an integral role in the exchange of materials (especially carbonates) in the atmosphere and oceans \citep{berner83}. On Mars, the cessation of plate tectonics may have been critical in the loss of a thick atmosphere and surface water \citep{kasting03}. Additionally, the differentiation of interior layers of a planet can affect energy and material transport. This differentiation is one of the primary requisites for liquid water layers in the moons of Jupiter and Saturn (see \S3.2). Finally, the composition and thickness of the outer crustal layers of a planet, such as an outer ice layer or crust, can contribute to the transport of nutrients from the surface to subsurface locales \citep{greenberg02}.

\subsubsection{Biological Feedback}
Once life forms, its impact on the environment will have profound affects on 
habitability. It is hypothesized that biological production of methane on early Earth may 
have been critical for maintaining a high surface temperature and exposed liquid water 
\citep{pavlov01}. Similarly, the rise of oxygen may have triggered global extinctions of 
anaerobic organisms; however, the presence of oxygen also produced an ozone layer, 
which provides a barrier against harmful UV radiation. Thus an oxygenic atmosphere opened the way for aerobic and multi-cellular life as well as land-based life, which does not have water to protect it from solar radiation (see \citet{kasting03} for a review of the evolution of the early Earth's atmosphere). 

\subsection{Characteristics Affecting Long-term Habitability}
Even if a planet's characteristics result in habitable conditions at a specific time, features of a planetÕs orbital evolution (e.g., inclination, eccentricity, obliquity) and interactions with its neighbors, the evolution of its host star as well as its climactic evolution (e.g., ice ages, carbon sinks) may result in temperature changes in time, with variability ranging from months to 10$^6$ years or greater. It remains unclear how severe a lapse in habitability must be to make the continued survival of life impossible, but evidence from mass extinction events on Earth suggest that once the distribution of simple life forms reaches a threshold it is difficult to extinguish completely.

\subsubsection{Dynamics of the Planetary System}
To maintain a consistent temperature, an Earth-like planet must continually maintain a nearly circular orbit or it will undergo extreme temperature changes.  In addition, the parameters of a planet's orbit can be affected by interactions with other planets in the system on both short and long timescales. 
Eccentric orbits increase the likelihood of collisions between planets; an eccentric 
giant planet can strongly inhibit stable orbits of other planets (see \S4.1). All the planets in our 
system have relatively circular orbits; however, many extrasolar planetary systems contain planets 
with highly eccentric orbits; the current distribution of extra-solar giant planet eccentricities is evenly distributed up to $\sim$0.7 (see the Extrasolar Planets Encyclopedia for current results; \citet{exocat}). Smaller bodies such as asteroids and comets can also affect 
the formation and survival of life: the composition of outer Solar System planetesimals makes them ideal for delivering water and organics during planet formation (see \S4.1); on the other 
hand, bombardment of a planet may kill developing biospheres, a process known as 
Òimpact frustrationÓ. 

\subsubsection{Host Star Mass and Evolution}
The radiative effects of the host star change with both stellar mass and the age of the star. More massive stars have shorter lifetimes than less massive stars, which may limit the probability of 
life arising on a planet around an early-type star. More massive stars also emit a larger fraction of 
their light in UV and X-ray wavelengths, which would have a detrimental impact on 
organic processes not protected by a thick atmosphere. Less massive stars than the Sun 
give off most of their light at longer wavelengths, potentially inhibiting or drastically modifying biological 
processes such as photosynthesis \citep{raven07}, and tidal locking at very small radii could cause atmospheric freeze-out on the dark side of the planet. Additionally, stars change their luminosity over their lifetime; the Sun was 70\% of its current luminosity when settled onto the main sequence \citep{gough81}. Therefore planets must have a sufficient feedback system and atmospheric volume to compensate for the changes in energy input.
   
\subsubsection{Presence of Satellites}
Earth is unusual with respect to the other rocky planets in our Solar System in that 
it has one large moon in a circular orbit. This could play a significant role in stabilizing 
Earth's rotation; the tilt of Earth towards the SunÑits ``obliquityÓÑis relatively stable 
over very long time periods. Without the Moon, Earth's obliquity could potentially vary 
drastically over million-year timescales, causing major surface temperature variations \citep{laskar93}. In 
addition, the formation of the Moon by a giant impact would have strongly affected the 
orbit and rotation of Earth, playing a major role in the final characteristics of the 
temperature and composition of the planet.  

\section{Assessing Habitability in our Solar System}
We can apply our understanding of the qualities that define habitable environments to the known planets and satellites in our own planetary system to gauge the potential for the origin and evolution of life, either ancient or extant. The current surface and subsurface conditions for planets interior to Earth (Mercury and Venus) suggest a very low probability for the origin and survival of life forms due to the very high surface temperatures ($\sim$700K) and very low water contents of both planets. The other terrestrial planet in the inner system, Mars, is a much better candidate, and recent results and upcoming investigations will be discussed in \S2.1. Beyond the Asteroid Belt, the potential for habitable environments is much less clear.  The best candidates appear to be the icy moons of Jupiter and Saturn, where evidence suggests a liquid water environment below an ice crust on both Europa and Enceladus; recent results will be discussed in \S2.2.

\subsection{Mars}
The enigmatic sister to Earth, Mars has long been considered our best candidate for detecting life on other planets; however, its promise as a habitable environment has fluctuated as we learn more about habitability in general and about the properties of Mars itself.

Mars is approximately 1/2 the size of Earth and has 1/10th the mass; it is also 1.52 times the distance to the Sun.  These factors contribute to a thin atmosphere (6/1000th as dense as Earth) and a mean surface temperature of -63\deg C (210 K).  However, the tilt of Mars' rotational axis (25\deg) is nearly equal to that of Earth, resulting in similar seasonal variations. The variation in solar insolation results in temperature fluctuations of $\sim$80\deg K between summer and winter, as well as polar caps that vary in size by a factor of 4.  It is clear from both the atmospheric conditions and high-resolution observations that significant amounts of stable liquid water, and therefore any macroscopic life that would rely on it, is not currently present on the surface of Mars.  Additionally, more exotic metabolism and cellular structures would be necessary to survive high doses of UV and X-ray radiation \citep{dartnell07,smith07} and oxidation reactions \citep{quinn05}; microbes on Earth do exist under these conditions \citep{houtkooper07}, but it is unclear if similar organisms could evolve on Mars. 

However, this does not preclude the presence of subsurface life.  Recent results from thermal imaging of the Martial surface suggest that water ice is stable as close as 20 cm from the surface \citep{banfield07}, and images of outflow gullies taken by the MGS demonstrate that a large amount of fluid, most likely water, was released from only tens of meters below the surface \citep{malin00}.  Additionally, investigations of surface conditions by the Mars rovers have found mineral sequences and sedimentology indicative of evaporation processes, as well as iron-rich mineral inclusions commonly known as ``blueberries" that commonly precipitate out of ground water  \citep{squyres04}. These results give credence to the idea that microbial life could be present in subsurface environments, and has sparked a resurgence of interest in both remote searches for atmospheric biomarkers as well as new in-situ experiments for local biomarkers below the surface.

\subsubsection{Remote Detection of Biomarkers}
Though it is difficult to probe the subsurface of Mars directly, we can sensitively test the atmospheric composition to search for evidence of byproducts of biological activity through orbiting spacecraft and ground-based observations from Earth.  However, discerning the difference between a biologically-produced atmospheric constituent and a gas produced abiotically is not trivial, and searches primarily focus on evidence of disequilibrium in the atmospheric chemistry compared with models.  

One of the primary signatures of disequilibrium is the coincidence of both oxidizing species (such as O$_2$) and reducing species (such as CH$_4$). The Martian atmosphere is dominated by CO$_2$ (95\%), and photodissociation easily produces trace amounts of other oxidizing species such as O$_2$ and OH.  Therefore the presence of a stable abundance of a reduced species in the Martian atmosphere would suggest ongoing production, either abiotic or biotic. \citet{summers02} suggests most biologically relevant gas species would have characteristic lifetimes of less than a year; methane, however, would survive for much longer ($\sim$300 years) and is thought to provide the best chance for detection.  Recent spectroscopic searches for methane in the Martian atmosphere have yielded ambiguous results, primarily due to the difficulty of searching for extremely weak features. \citet{formisano04} reported a marginal detection of 10$\pm$5 ppb methane using the Planetary Fourier Spectrometer onboard the Mars Express orbiter, but the severe limitations in spectral resolution and sensitivity have cast doubt on these results.  \citet{krasnopolsky04} published a similar result (10$\pm$3 ppb) using ground-based NIR spectroscopy, but this study was also hampered by instrumental uncertainties and the difficulty of removing the terrestrial methane signature; a more recent search produced only upper limits \citep{krasnopolsky07}.  Considering the uncertainties in these results a judgement on the presence of methane in the Martian atmosphere would clearly be premature, and we must wait for improved observing resources and/or analysis; several sensitive campaigns are currently being undertaken (i.e. \citet{mumma07}).

Interpreting these and any future detections or non-detections is further complicated by the possibility of abundance variations due to seasonal or local release. Even if methane were to be securely detected, its provenance would be unclear; there are a variety of abiotic production mechanisms for methane that have been proposed, ranging from cometary delivery \citep{kress04} to the photolysis of water and CO in the atmosphere \citep{barnun06} and low-temperature alteration of basalts (``serpentinization") \citep{oze05}. Future studies must therefore focus on discerning the observable signatures of various production pathways before we can elucidate the true origin of any detected species.

\subsubsection{In-situ Detection of Biomarkers}
Finding evidence of life from the surface of Mars is no less fraught with pitfalls than the remote detection of life.  The first experiments to test for Martian life on the surface, conducted on the Viking Lander 1 in 1976, led to more questions than they answered (see \citet{schuerger07} for a recent review). Three different biology experiments, two gas release experiments and a pyrolytic release experiment, yielded what were considered positive results but were later attributed to abiotic processes.  The most damning case against biology came from the results of the gas chromatograph-mass spectrometer which failed to find any trace of organic material, the basic building blocks of all life on Earth. The validity of these conclusions is still under debate (i.e. \citet{benner00,houtkooper07}), but they continue to inform designs for future searches for evidence of life on or below the surface of Mars.

The first upcoming Mars lander mission to test for evidence of surface habitability will be the NASA Phoenix Lander, currently set to touch down in May of 2008.  Phoenix will not conduct any experiments to directly detect extant or extinct life, but it will further characterize the organic material and level of oxidation of the Martian soil with much greater sensitivity than the Viking experiments using a wet soil chemistry experiment (MECA) and a gas-release mass spectrometer experiment (TEGA) \citep{shotwell05}.  More importantly, Phoenix is equipped with a robotic arm that can dig up to 0.5m below the surface.  Tests of subsurface material will measure the hydration level as well as the degree of oxidation and photolysis of organic material, adding significantly to our understanding of the potential for extant subsurface life. 

The first post-Phoenix missions to test for life will be the NASA Mars Science Laboratory (MSL) and the ESA ExoMars mission, both rovers carrying a wide variety of instruments.  Both missions will extend Phoenix's investigations of the organic inventory in both the sensitivity and the scope of the experiments performed. The primary life-detection experiment planned for the MSL, expected to launch is 2009, is known as SAM (Sample Analysis at Mars). SAM will be a combination of a gas chromatograph and both a mass spectrometer and a tunable laser spectrometer \citep{mahaffy07}.  The gas chromatograph will reach higher temperatures than previous experiments ($\sim$1100\deg C), sufficient to observe both volatile species indicative of extant biology and moderately refractory species indicative of past life, and the coupled GC-MS will be sensitive down to a fraction of a picomole of organic material. ExoMars, expected to launch in 2013, will aim to include a Raman-LIBS (laser-induced breakdown spectrometer) for organic analysis, an oxidant sensor, a ``life marker chip" which would use antibody reactions to detect specific molecules with extreme sensitivity, as well as a drill to reach $\sim$2m below the surface \citep{exomars06}. 
These two missions are expected to fully characterize the astrobiological properties of the surface and immediate sub-surface of Mars; future improvements would be left to either a sample return mission or a manned mission.

\subsection{Europa \& Enceladus}
In the outer Solar System, the low solar insolation level results in surface temperatures far below the freezing point of water.  Additionally, volatile inventories for most bodies are high compared with the inner Solar System, resulting in low bulk densities and icy surfaces (first confirmed through Voyager imaging; \citet{smith79}) for many of the satellites of the giant planets. In addition, tidal stresses related to orbital resonances were shown to be a potential heat source for generating a liquid ocean below the ice crust of at least one Galilean moon, Europa \citep{cassen79,squyres83}. More recently, direct evidence of liquid water from an icy moon was confirmed through Cassini observations of a plume of water-rich material flowing from the south pole of Enceladus, a moon of Saturn \citep{porco06}. Investigations of these potentially habitable environments are still in their infancy, but they offer a tantalizing new option for finding life beyond Earth.
 
Europa has a radius of 1560 km, close to that of the Earth's moon. In 1998 radio Doppler data from the Galileo spacecraft demonstrated a high probability for a differentiated internal structure and a thick outer ice shell on Europa \citep{anderson98}, and the lack of significant cratering and other signatures of resurfacing \citep{zahnle98} suggest an active geologic history possibly aided by a liquid water layer below the ice. This hypothesis was further supported by magnetometer results from Galileo requiring a near-surface global conducting layer, most plausibly in the form of a saline water ocean \citep{kivelson00}. Current analysis of topography data suggests an ice shell thickness of approximately 15 - 25 km \citep{nimmo03}, while the magnetometer data suggests an ice shell thickness of less than 15km \citep{hand07}. Beyond liquid water, the two primary uncertainties with regard to habitability of a subsurface ocean are the availability of biologically-important compounds and the free energy to assemble and maintain them.   Volcanic and tectonic activity as well as hydrothermal vents may contribute to the ocean mineralogy \citep{reynolds83}; additionally, cometary impacts can deliver small amounts of biogenic elements \citep{pierazzo02}. However, without knowledge of the water chemistry or composition and structure of the sub-ocean mantle it is difficult to estimate the contribution of material or heat through such processes.  

Enceladus, only 500 km in diameter, was not considered to be a viable astrobiological target due to a lack of internal heating until Cassini revealed an outburst of material emanating from a hot spot near the moon's south pole in 2005.  Images of the south pole suggest a recent resurfacing, with ``tiger striped" ridges indicative of tectonic activity; temperatures measured for the ridges by the Visual and Infrared Mapping Spectrometer are in the range of 140K \citep{brown06}. During Cassini's passage through the plume, measurements from the Ion and Neutral Mass Spectrometer showed the composition to be mostly water, with traces of methane and N$_2$ \citep{waite06}; the chemical composition shows evidence of production in a hot (T$\sim$500K) catalytic environment \citep{matson07}. The outgassing may result either from shallow liquid water reservoirs \citep{porco06} or from clathrate disruption \citep{kieffer06}; however, it is unclear how the heating necessary to produce the surface features would be generated and transported in the first place. Tidal shearing presents a viable option for heat transport, but would most likely require a global sub-surface ocean \citep{nimmo07}; the flattened shape of the south pole also suggests warming due to a local subsurface sea \citep{collins07}, but alternative models suggest a low-density ice flow \citep{nimmo06}. However, the effects of both past and present tidal heating and radiogenic heating may be insufficient to produce the required heat flux under realistic conditions \citep{meyer07,schubert07}; without a clear understanding of the temporal nature of the heating processes it is difficult to assess the validity of specific models. Additional data on the gravitational anomaly from future Cassini fly-bys will help to resolve some of the ambiguities.

The next step in exploring the habitability of both Europa and Enceladus would be a thorough investigation of the surface ice characteristics through instruments for ground-penetrating radar, altimetry, high-resolution imagery and near-infrared spectroscopy to further discern the surface and subsurface chemistry \citep{chyba02}. For Europa, ESA is considering an orbiter mission known as the Jovian Minisat Explorer, while NASA has considered both a Europa Orbiter as well as the Jupiter Icy Moons Orbiter; however, both concepts are still highly uncertain and will most likely be reconsidered due to the new Enceladus results.   

\section{Finding Habitable and Inhabited Planets Around Other Stars}
For almost 500 years, from the time of Copernicus' discovery of the heliocentric nature of our Solar System until the discovery of the first extra-solar planet in 1995 \citep{mayor95}, humankind attempted to understand the origin of our living planet and the rest of the celestial bodies through the narrow lens of our single planetary system.  We therefore developed complex theories on the formation of planets that naturally result in a high probability of producing a Solar System just like our own.  However, it is not altogether surprising that with the discovery of a second planetary system, these theories were abruptly turned upside down. With more than 250 giant planets now known to orbit main-sequence stars (\citet{butler06}; see \citet{exocat} for recent results), theories on the planet formation and evolution developed for our own planetary system must be re-examined.

\subsection{Understanding Habitable Planet Formation}
To predict where to look for habitable planets, and what types of characteristics the planets we find will have, we must understand the evolution of the initial gas-and-dust-rich protoplanetary environments in which planets grow and the forces that shape the formation and evolution of these planets into habitable worlds.

The primary factors that define whether habitable planets can form and remain stable for biologically significant timescales are:
\begin{itemize}
\item A mass density near the Habitable Zone sufficient to form planets capable of sustaining an atmosphere and geologic activity
\item A volatile contribution (most importantly water) sufficient to sustain the origin and evolution of life
\item A sufficiently stable planetary system such that any orbital variations do not result in long periods of inhabitability
\end{itemize}
The first two factors are determined primarily by the initial conditions of the protoplanetary nebula and the subsequent evolution of raw material into a young planetary system as solid bodies accrete and the gaseous disk dissipates. The stability of a planetary system is primarily determined by the post-accretion dynamics of the system when planets begin to interact and scatter each other within (and out of) the system.  Our own Solar System clearly passed all three tests:  sufficient mass and volatile material was available to create at least one planet capable of harboring life, and our planetary system was stable enough that Earth was able to remain on an almost circular orbit for billions of years. The question remains as to whether our Solar System represents the norm, or just a rare aberration.

\subsubsection{Standard Planet Formation Theory}
Standard theories of the evolution of planetary cores suggest that solid material in a circumstellar disk will proceed through various accretionary stages culminating in the final architecture of a stable planetary configuration. Coagulation of the inceptive dust particles occurs through collisional sticking to produce meter-sized objects on timescales of $10^4$ years \citep{lissauer93}.  Once the largest bodies reach $\sim$1 km in size, their gravitational cross-section becomes larger than their geometric cross-section and they begin the phase known as ``runaway growth" \citep{greenberg78,wetherill89}, eventually leading to a binomial distribution of ``embryos" ($M\sim0.1$ M$_{\oplus}$) and ``planetesimals" ($M<10^{-3}$ M$_{\oplus}$) after approximately $10^{6-7}$ years. The final ``chaotic phase" of planet growth proceeds through scattering and collisions between the large protoplanets and final clearing of the remaining planetesimals to produce a stable planetary system after more than $10^8$ years \citep{wetherill96,kenyon06}.

According to traditional theories of planet formation, these processes occur primarily in localized regions, with little mass transport between the inner and outer disk.  The initial composition of solids in a circumstellar disk is expected to follow a basic condensation sequence \citep{grossman72}; therefore, planets formed beyond the current ``snow line" at approximately 2.7 AU, where the enhanced mass density due to volatile freeze-out leads to more massive embryos \citep{stevenson88}, would be able to reach a critical core mass and initiate run-away gas accretion to form gas giants \citep{pollack96}. Rocky planets formed inside the snow line would be relatively water-poor; water-rich material could then be delivered through late-stage cometary impacts \citep{owen95} or inward scattering of icy asteroids \citep{morbidelli00}. Planets would remain on almost circular orbits due to the relatively quiescent formation process, with the only major rearrangements occurring due to impacts between embryos such as the Moon-forming impact \citep{hartmann75}.

\subsubsection{New Results from Extrasolar Planets and Planet Formation Studies}
The discoveries of extrasolar planetary systems, new results on evolutionary timescales for our own Solar System and other protoplanetary disks, and the advent of sophisticated hydrodynamic and N-body investigations of protoplanetary disks have shed new light on the processes at work in planet formation, while also leading to new conundrums. Though radial velocity (RV) searches are only complete out to approximately 3 AU \citep{butler06}, there is already clearly a pile-up of planets at very small semi-major axes: 44\% of RV-detected extrasolar planets have orbital radii less than 0.3 AU, and 18\% are located within 0.05 AU of the central star. Recent models of gas-rich accretion disks support the theory that giant planets form beyond the snow line and migrate inwards in a process known as ``Type II migration" \citep{papaloizou84,lin96}; additional migration processes have also been shown to function on smaller bodies embedded in gaseous disk (``Type I migration" for Earth-sized bodies \citep{ward97,masset06} and drag forces for planetesimals \citep{ciesla06}). These processes will have a profound affect on mass transport during planet formation, and when combined with models that demonstrate the effects of variable condensation fronts throughout the disk lifetime \citep{garaud07,kennedy07} these results transform our understanding of the transport and delivery of water and volatiles to the terrestrial planets. 
 
Additionally, both the orbital eccentricities and masses of known extrasolar planets with semi-major axes between 0.1 AU and 3 AU are remarkably evenly distributed up to $e=0.7$ and $M=4$ M$_J$ \citep{butler06,exocat}. The upper mass limit is mostly likely a result of initial disk mass; however, there are a number of plausible mechanisms for exciting eccentricities such as planet-disk interactions \citep{goldreich03,kley06,dangelo06}, perturbations from stellar companions or field stars \citep{zakamska04,takeda05} and planet-planet scattering \citep{marzari02,ford07}, and it is unclear what role (if any) each process plays in each system.

These new results present challenges for both explaining our own Solar System's architecture as well as  predicting the characteristics of terrestrial planets in other planetary systems.  It is still unclear whether our system suffered migration of a giant planet or rocky cores during the earliest stages of formation, though the mass and semi-major axis ratios of Jupiter and Saturn could have acted to stop or reverse Type II migration \citep{masset01,morbidelli07}. The role of inward migration of water-rich material on the volatile content of the inner system has not been adequately addressed, though modeling improvements have been made \citep{raymond07}; further work is also necessary to incorporate the effects of fragmentation and heating processes.

Initial investigations of terrestrial planet formation in extrasolar planetary systems are encouraging. Models of planetary systems that experience migration of a giant planet to the inner system have demonstrated that habitable planets can form and survive in these systems \citep{mandell03,fogg05}; the planets formed are usually more massive and water-rich than those in our own Solar System \citep{raymond06b,mandell07}.  Systems which experience the scattering of a giant planet into an eccentric orbit near the Habitable Zone can clear out planetesimals \citep{veras05}, but terrestrial planet formation in a system with giant planets in stable orbits beyond 2.5 AU will be uninhibited \citep{raymond06}.  However, many recent advances in our understanding of the evolution of circumstellar disks are not yet fully incorporated into these models, and results must be regarded as preliminary.

\subsection{Current and Future Searches for Habitable / Inhabited Planets}
Current extrasolar planet detection techniques are not yet sensitive enough to detect Earth-like planets around Sun-like stars in the Habitable Zone; the current detection limit for solar-type stars is approximately 10 M$_{\oplus}$, and that is only for close-in planets in multi-planet systems \citep{exocat}. However, low-mass M stars present a much better chance for habitable planet detection: the lower stellar mass allows for detection of smaller planets, and the lower luminosity results in a Habitable Zone close to the parent star. The only currently known M-star planetary system which may be habitable is GL581, with a 5 M$_{\oplus}$ planet at 0.073 AU and an 8 M$_{\oplus}$ planet at 0.25 AU \citep{udry07}. The two planets straddle the traditional Habitable Zone defined solely by the stellar insolation; however, atmospheric circulation models suggest a thick atmosphere on the tidally-locked outer planet could increase the temperature to within the habitable range \citep{vonbloh07,selsis07}.

Discoveries of habitable planets more like Earth will most likely occur through future space missions such as the Kepler transit search \citep{basri05} (estimated launch date of 2009) and the SIM PlanetQuest astrometry mission \citep{catanzarite06} (estimate launch date of 2016). The 1-meter Kepler telescope will monitor $\sim$100,000 main-sequence stars (m$_v = 9 - 14$) continuously for 4 years, with precision sufficient to detect Earth-mass planets at 1 AU around solar-mass stars with m$_v$=12. Kepler has the potential to find hundreds of Earth-like planets, but the host stars will be beyond the range of most follow-up ground-based techniques. SIM PlanetQuest will most likely be a visible-wavelenth interferometer with two 0.3m telescopes separated by a 9-meter baseline, capable of detecting proper motions with 1 $\mu$as precision. Though its detection numbers for Earth-like planets ($\sim$10) would be much smaller than Kepler, it could observe nearby stars with and without detected planets, furthering the characterization of nearby planetary systems currently being observed by ground-based campaigns.

The search for life on these planets will have to wait until the launch of the Terrestrial Planet Finder (NASA) and/or Darwin (ESA) missions, which will seek to simultaneously image nearby Earth-like planets and analyze their atmospheres for biomarkers using low-resolution visible or near-infrared spectroscopy.  Both missions are in the early design phases, but considerable work has been done on predicting the spectral signatures of planets with a variety of surface features (see \citet{kaltenegger07} for a recent review). Primary molecular biomarkers focus on disequilibrium chemistry between oxygen species and reducing organics such as methane, with studies showing variations with cloud cover, surface ice and vegetation fraction. However, the current lack of constraints on potential variations in the characteristics of planetary surfaces and atmospheres, variations in biological evolution under different conditions, and firm plans for instrumentation make constraining potential biosignatures difficult.

Though we cannot currently search for biosignatures, the search for abiotic signs of life continues through various SETI (Search for Extraterrestrial Intelligence) programs. Due to rapid advances in signal processing sophistication and sensitivity, searches now include all regions of the temporal (i.e. pulses versus continuous) and frequency domains (from optical to radio wavelengths). A huge leap in coverage of the GHz regime will be accomplished with the Allen Telescope Array and the future Square Kilometer Array (SKA), with the added benefit of flexible resource allocation based on target availability and priority \citep{tarter04}. The ability to target large numbers of stars, combined with improved constraints based on planet detection missions, will vastly improve the SETI search efficiency.  It may be a fruitless search, but one confirmed detection would alter our perception of our place in the universe forever.

\section{Conclusion}
While this review is in no way an exhaustive account of the state of astrobiology-related research, it profiles the most recent results in the search for life beyond the confines of our planet, both within our own planetary system and beyond.  The prospects for detecting evidence of life elsewhere in the next few decades, or at least constraining the planetary environments in which life cannot evolve, are bright: within 10 years we will explore the immediate subsurface of Mars, conduct sensitive searches for Earth-like planets around other stars, and drastically improve our understanding of the characteristics of these potentially habitable planetary systems. In the subsequent decade we will hopefully begin further exploration of the icy moons in the outer Solar System, discovery and characterize nearby Earth-like planets, and possibly detect life on these planets through atmospheric biomarkers.  Though we currently appear as a single oasis in a vast empty desert, by the year 2025 the Earth may be only one of many examples of locales amenable for the origin and evolution of life.  

\acknowledgements 
I would like to thank the Astronomy Department at the University of Texas at Austin for giving me the opportunity to participate in the Bash Symposium, and G. Villanueva for stimulating
conversations during the preparation of this manuscript. Support for this work was provided by the NASA through the NASA Post-doctoral Program.

\bibliographystyle{apj}

\end{document}